\documentclass{iaus}
\usepackage{graphicx}

\title[RDI and Triggered Star Formation] 
{Radiation Driven Implosion and Triggered Star Formation}

\author[T. G. Bisbas, A. P. Whitworth, R. W\"unsch, D. A. Hubber, \& S. Walch]
{T. G. Bisbas$^1$,
A. P. Whitworth$^2$,
R. W\"unsch$^1$,
D. A. Hubber$^3$,
\and S. Walch$^2$}

\affiliation{$^1$Astronomical Institute, Academy of Sciences of the Czech Republic, Bo\v{c}n\'{i} II 1401, 141 31 Prague, Czech Republic. Email: {\tt t.bisbas@astro.cf.ac.uk}\\[\affilskip]
$^2$School of Physics and Astronomy, Cardiff University, Queens Buildings, The Parade, Cardiff, CF24 3AA, United Kingdom\\[\affilskip]
$^3$Department of Physics and Astronomy, University of Sheffield, Hicks Building, Hounsfield Road, Sheffield S3 7RH, United Kingdom}

\pubyear{2010}
\volume{270}  
\jname{Computational Star Formation}
\editors{eds. Alves, Elmegreen, Girart, Trimble}

\newcommand{\HIIR}{H{\sc ii} region}

\begin{document}

\maketitle

\begin{abstract}
We present simulations of initially stable isothermal clouds exposed to ionizing radiation from a discrete external source, and identify the conditions that lead to radiatively driven implosion and star formation. We use the Smoothed Particle Hydrodynamics code SEREN (\cite[Hubber et al. 2010]{Hubber2010}) and the HEALPix-based photoionization algorithm described in \cite[Bisbas et al. (2009)]{Bisbas2009}. We find that the incident ionizing flux is the critical parameter determining the evolution: high fluxes simply disperse the cloud, whereas low fluxes trigger star formation. We find a clear connection between the intensity of the incident flux and the parameters of star formation.
\end{abstract}

\firstsection 
\section{Introduction}

When an expanding {\HIIR} overruns a pre-existing cloud, it compresses it by driving an ionization front and a shock wave into it (\cite[Sandford et al. 1982]{Sandford 1982}; \cite[Bertoldi 1989]{Bertoldi1989}; \cite[Lefloch \& Lazareff 1994]{LL94}). The inner parts may become gravitationally unstable and collapse to form new stars. This mechanism is known as Radiation Driven Implosion (RDI). Observations (\cite[Lefloch \& Lazareff 1995]{LL95}; \cite[Lefloch et al. 1997]{Lefloch1997}; \cite[Sugitani et al. 1999]{Sugitani1999}, \cite[2000]{Sugitani2000}; \cite[Ikeda et al. 2008]{Ikeda2008}; \cite[Morgan et al. 2008]{Morgan2008}; \cite[Chahuan et al. 2009]{Chahuan2009}) strongly support a connection between the RDI mechanism and the formation of Young Stellar Objects (YSO). Simulations of the interaction of the ultraviolet ionizing radiation with self-gravitating clouds have been presented by various authors (\cite[Kessel-Deynet \& Burkert 2003]{Kessel2003}; \cite[Gritschneder et al. 2009]{Grit2009}; \cite[Miao et al. 2009]{Miao2009}). However, no model can explain \emph{where} star formation takes place (in the core or at the periphery) or \emph{when} (during the maximum compression phase or earlier -- \cite[Deharverg et al. 2005]{Deharverg2005}).

In this paper we perform a set of 75 simulations of clouds exposed to ionizing radiation. The aim of this work is to answer questions of whether the incident ionizing flux is able to trigger the formation of new stars or not, and how the process and the properties of this star formation are connected with the intensity of the incident flux. In Section 2 we give a brief description of the numerical treatment and the initial conditions we use. In Section 4 we discuss the results of our simulations. We summarize in Section 5.

\section{Numerical Treatment and Initial Conditions}

We use the Smoothed Particle Hydrodynamics (SPH) code SEREN\footnote{http://www.astro.group.shef.ac.uk/seren}, fully described in \cite[Hubber et al. (2010)]{Hubber2010}, with an ionization routine (Bisbas et al. 2009) based on the HEALPix\footnote{http://healpix.jpl.nasa.gov} sphere tesselation code (\cite[G{\'o}rski et al. 2005]{Gorski2005}). The temperature of the neutral gas at density $\rho$ is $T_{_{\rm N}}(\rho)=T_{_{\rm ISO}}\left\{1+\left(\rho/\rho_{_{\rm CRIT}}\right)^{\gamma-1}\right\}$, where $T_{_{\rm ISO}}=10\,{\rm K}$, $\rho_{_{\rm CRIT}}=10^{-13}\,{\rm g}\,{\rm cm}^{-3}$ and $\gamma=5/3$ is the ratio of specific heats. The temperature of the ionized gas is taken to be $T_{\rm i}=10^4\,{\rm K}$, except in the transition zone between the two extremes, where it changes smoothly from $T_{\rm i}$ to $T_{_{\rm N}}$ (see \cite[Bisbas et al. 2009]{Bisbas2009}). We include sink particles (\cite[Bate et al. 1995]{Bate1995}) with radii $R_{_{\rm SINK}}=2.5\,{\rm AU}$ created if $\rho>\rho_{_{\rm SINK}}=10^{-11}\,{\rm g}\,{\rm cm}^{-3}$.

Our clouds are stable Bonnor-Ebert spheres (\cite[Bonnor 1956]{Bonnor1956}; \cite[Ebert 1957]{Ebert1957}; heareafter `BES') with dimensionless cut-off radii $\xi_{_{\rm B}}=4,5,6$ and with masses $M=2,5,10\,{\rm M}_{\odot}$. The particle resolution we use is $5\times10^4$ SPH particles per solar mass (c.f. \cite[Hubber et al. 2006]{Hubber2006}). We place the BESs at distance $D=10R$, where $R$ is the radius of the cloud (in pc), in order to keep constant the divergence of the incident flux and as parallel as possible. We use a single source emitting Lyman-$\alpha$ photons. We run simulations with a wide range of emission rates ${\dot{\cal{N}}}_{_{\rm LyC}}=10^x\,{\rm s}^{-1}$, where $x=48,\,48.5,...52$.

\section{Results}

\begin{figure}[b]
\begin{center}
 \includegraphics[width=0.36\textwidth]{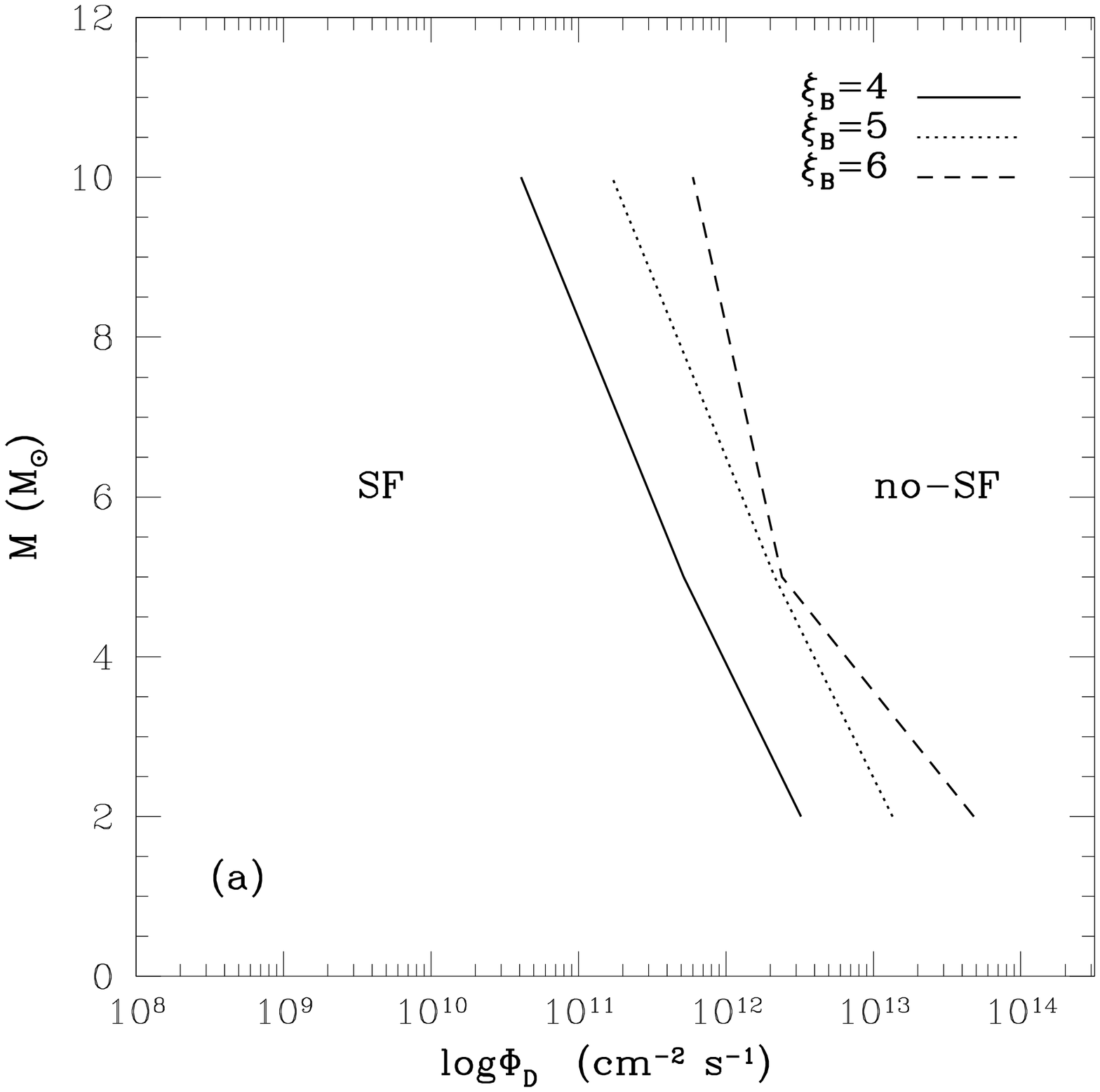} 
 \includegraphics[width=0.36\textwidth]{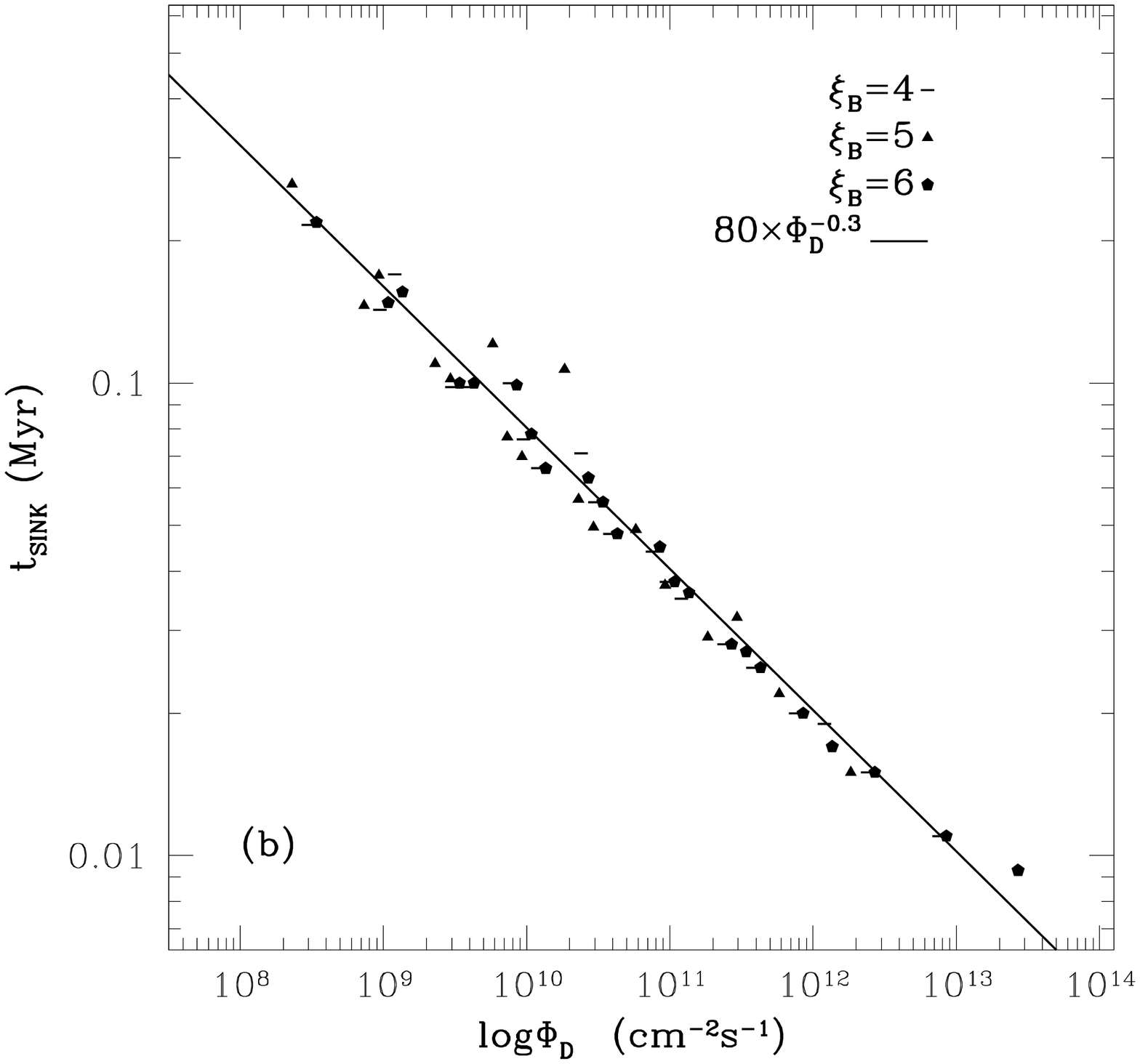} 
 \includegraphics[width=0.36\textwidth]{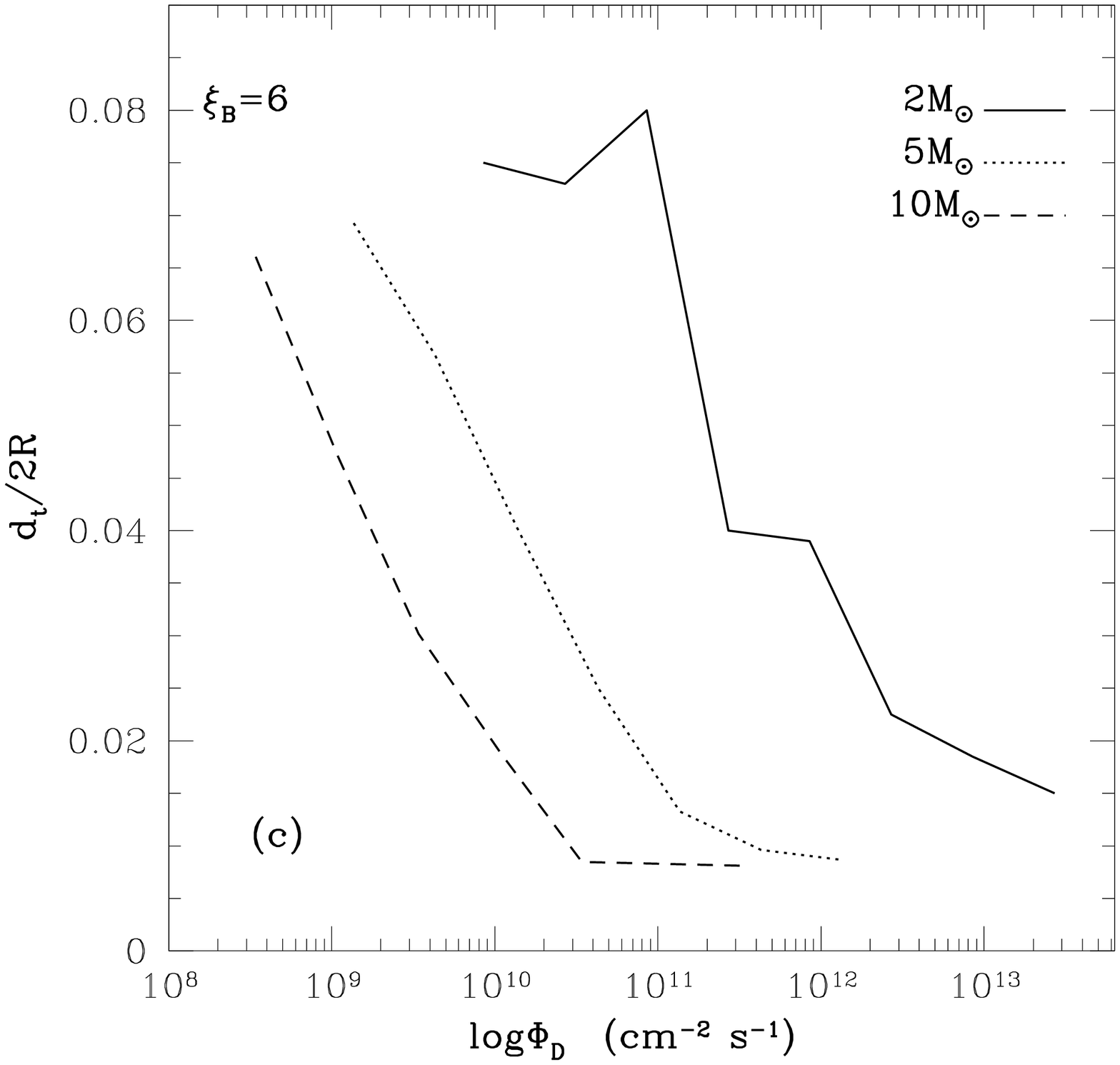}
 \includegraphics[width=0.36\textwidth]{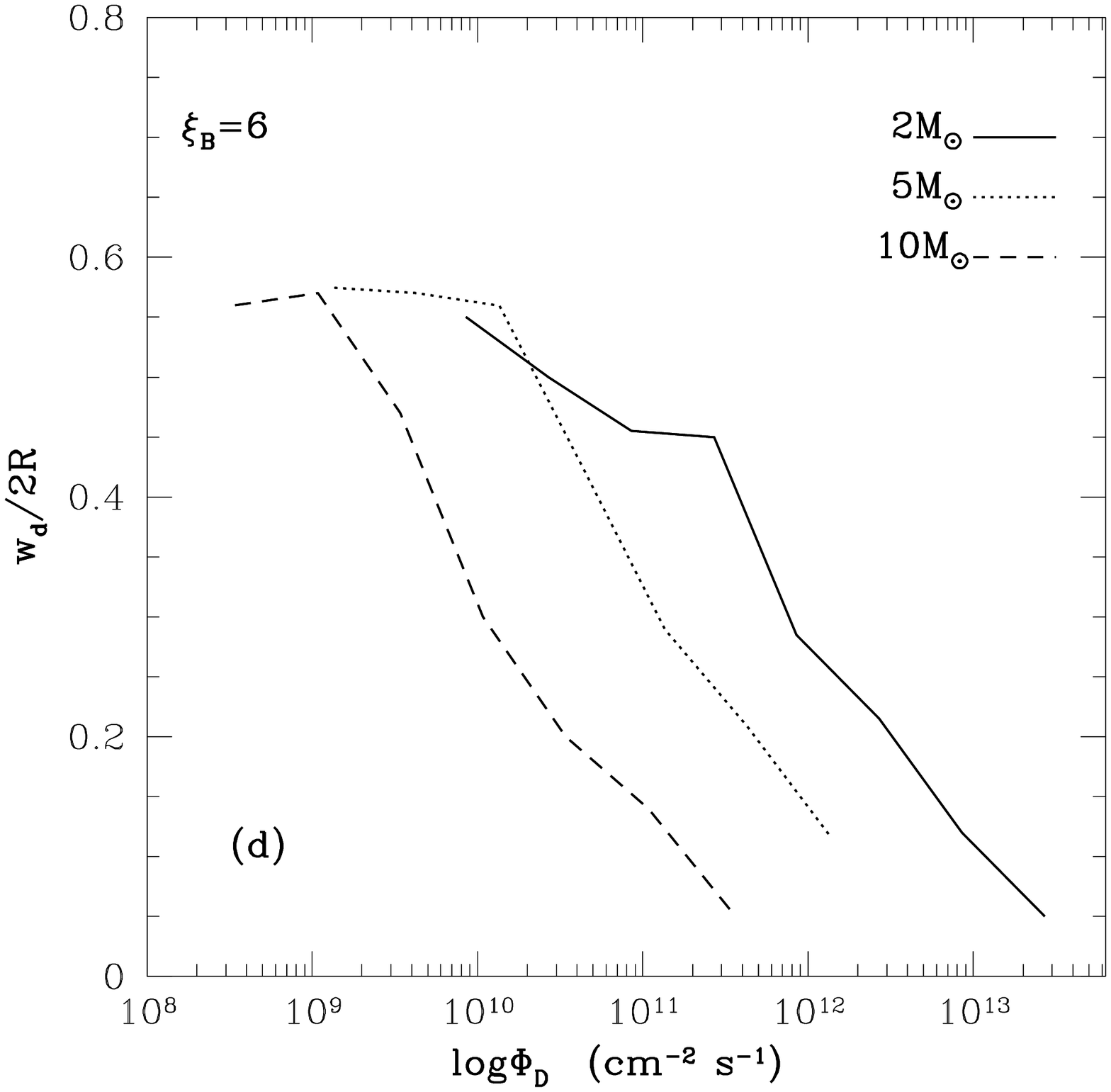}
 \caption{ (a) The flux-mass semi-logarithmic diagram where we define areas where stars are formed (SF) and areas where stars are not formed (no-SF) depending on the dimensionless radius $\xi_{_{\rm B}}$ of a BES. (b) Logarithmic diagram of the incident flux versus the age of the cloud, $t_{_{\rm SINK}}$, when star formation occurs. The power law we propose (solid line) fits very well with our simulations ($t_{_{\rm SINK}}$ is in Myr and $\Phi_{_{\rm D}}$ is in ${\rm cm}^{-2}\,{\rm s}^{-1}$). (c) Star formation occurs at the periphery with increasing flux. (d) Star formation occurs during maximum compression with increasing flux.}
   \label{fig.all}
\end{center}
\end{figure}

In Fig.\ref{fig.all}a we present a semi-logarithmic diagram where we correlate the intensity of the incident ionizing flux with the initial mass of each BES. The lines define subsets of parameter space where models either show star formation (left) or not (right), with accuracy 0.25 dex. It can been seen that as the mass of the BES decreases (and as a result $\xi_{_{\rm B}}$ increases) the clouds appear to survive longer in higher fluxes. This is because for a given $\xi_{_{\rm B}}$, the density $\rho_{\rm c}$ at the centre of each BES increases with decreasing $M$. 

We also find that the Str{\o}mgren radius at the end of the $R$-type expansion determines whether stars are formed or not; if the ionization front has not overrun the central core of the BES, then the incident flux will trigger star formation during the $D$-type expansion of the {\HIIR}.

The time, $t_{_{\rm SINK}}$, between the beginning of the $D$-type expansion and the first sink creation (beginning of star formation) increases with decreasing ionizing flux. This finding is in agreement with simulations of the RDI performed by \cite[Gritschneder et al. (2009)]{Grit2009}. Figure \ref{fig.all}b is a logarithmic diagram where we plot the values of $t_{_{\rm SINK}}$ versus the incident flux $\Phi_{_{D}}$. Remarkably, the age of the cloud when star formation occurs does not depend on its properties (i.e. the initial mass, $M$, or $\xi_{_{\rm B}}$). Results from our simulations can be described with a power law of the form $t_{_{\rm SINK}}=80\times\Phi_{_{\rm D}}^{-0.3}$ ($t_{_{\rm SINK}}$ in Myr, $\Phi_{_{\rm D}}$ in ${\rm cm}^{-2}{\rm s}^{-1}$).

\begin{figure}[h]
\begin{center}
 \includegraphics[width=0.72\textwidth]{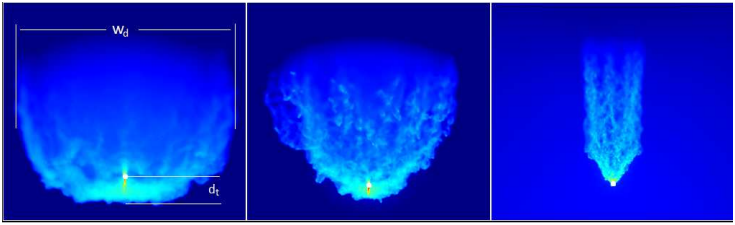}
\caption{ Column density plots of a BES with $M=10\,{\rm M}_{\odot}$ and $\xi_{_{\rm B}}=6$ at $t_{_{\rm SINK}}$ when it is exposed to three different intensities of flux (flux increases from left to right). The white dots are sink particles. In the left plot we draw the values of $d_{\rm t}$ and $w_{\rm d}$.}
\label{fig.morphology}
\end{center}
\end{figure}

Figure \ref{fig.morphology} shows column density plots of a BES with $M=10{\rm M}_{\odot}$ and with $\xi_{_{\rm B}}=6$ at $t_{_{\rm SINK}}$ for different fluxes. A common feature in all our simulations is that stars form close to the symmetry axis joining the centre of the cloud to the exciting star. This is in an agreement with observations by \cite[Sugitani et al. (1999)]{Sugitani1999}. The distance $d_{\rm t}$ between the first sink particle and the ionization front is a function of the ionizing flux and the BES parameters (see Fig.\ref{fig.all}c where we plot $d_{\rm t}/2R$ for all BESs with $\xi_{_{\rm B}}=6$). We find that for low fluxes stars tend to form in the innermost part of the filamentary structure, whereas for high fluxes stars tend to form at the periphery of the cloud. Similar results are found also with $\xi_{_{\rm B}}=4$ and $\xi_{_{\rm B}}=5$. 

Figure \ref{fig.morphology} shows that the lateral compression, $w_{\rm d}$, of the BESs at the beginning of star formation is connected to the intensity of the incident flux. We see that for low fluxes, $w_{\rm d}$ is quite high and the cloud has a {\bf U}-shape structure, whereas for high fluxes $w_{\rm d}$ is small and the cloud has a {\rm V}-shape stucture. In Fig.\ref{fig.all}d we plot $w_{\rm d}/2R$ for all BESs with $\xi_{_{\rm B}}=6$ and we find that stars tend to form during maximum compression once the incident flux is increased. Similar results are found also for the rest of the clumps.

\section{Conclusions}

We present simulations of Radiation Driven Implosion in stable clouds represented by Bonnor-Ebert spheres. We performed 75 simulations with clouds of different masses, different dimensionless radii, and with a wide range of incident fluxes. In general we find a connection between the incident ionizing flux and the resultant star formation efficiency.

We introduce a semi-logarithmic diagram (flux-mass diagram) where we correlate the intensity of the incident flux and the initial mass of each BES, and we define zones of Star Formation and no-Star Formation. We find that if the Str{\o}mgren radius at the end of the $R$-type expansion has not overrun the central core of the BES, the ionizing radiation will trigger star formation. The age of the cloud when star formation occurs increases with decreasing flux, and it does not depend on the properties of the BES. A power-law of the form $t_{_{\rm SINK}}=80\times\Phi_{_{\rm D}}^{-0.3}$ fits very well with the results of our models. Finally, as the incident flux increases, stars tend to form closer to the periphery of the cloud and during its maximum compression phase. 

\noindent\textbf{Acknowledgments:} TGB and RW acknowledge support from the project LC06014-Centre for Theoretical Astrophysics of the Ministry of Education, Youth and Sports of the Czech Republic. APW and SW gratefully acknowledge the support of the Marie Curie Research Training Network {\sc CONSTELLATION} (Ref. MRTN-CT-2006-035890). DAH is funded by a Leverhulme Trust Research Project Grand (F/00 118/BJ). The computations in this work were carried out on Merlin Supercomputer of Cardiff University. The data analysis and the column density plot were made using the SPLASH visualization code (\cite[Price 2007]{Price2007}).


\begin{thebibliography}{}

\bibitem[Barnes \& Hut(1986)]{BH1986} Barnes, J., \& Hut, P.\ 1986, \nat, 324, 446 

\bibitem[Bate et al.(1995)]{Bate1995} Bate, M.~R., Bonnell, I.~A., \& Price, N.~M.\ 1995, \mnras, 277, 362 

\bibitem[Bertoldi(1989)]{Bertoldi1989} Bertoldi, F.\ 1989, \apj, 346, 735

\bibitem[Bisbas et al. (2009)]{Bisbas2009} Bisbas, T.~G., W{\"u}nsch, R., Whitworth, A.~P., \& Hubber, D.~A.\ 2009, \aap, 497, 649 

\bibitem[Bonnor(1956)]{Bonnor1956} Bonnor, W.~B.\ 1956, \mnras, 116, 351

\bibitem[Chauhan et al.(2009)]{Chauhan2009} Chauhan, N., Pandey, A.~K., Ogura, K., Ojha, D.~K., Bhatt, B.~C., Ghosh, S.~K., \& Rawat, P.~S.\ 2009, \mnras, 396, 964

\bibitem[Ebert(1957)]{Ebert1957} Ebert, R.\ 1957, Zeitschrift fur Astrophysik, 42, 263

\bibitem[Deharveng et al.(2005)]{Deharveng2005} Deharveng, L., Zavagno, A., \& Caplan, J.\ 2005, \aap, 433, 565

\bibitem[G{\'o}rski et al.(2005)]{Gorski2005} G{\'o}rski, K.~M., Hivon, E., Banday, A.~J., Wandelt, B.~D., Hansen, F.~K., Reinecke, M., \& Bartelmann, M.\ 2005, \apj, 622, 759

\bibitem[Gritschneder et al.(2009)]{Grit2009} Gritschneder, M., Naab, T., Burkert, A., Walch, S., Heitsch, F., \& Wetzstein, M.\ 2009, \mnras, 393, 21

\bibitem[Hubber et al.(2010)]{Hubber2010} Hubber, D.~A., Batty, C.~P., McLeod, A., \& Whitworth, A.~P.,\ 2010, \aap, submitted

\bibitem[Hubber et al.(2006)]{Hubber2006} Hubber, D.~A., Goodwin, S.~P., \& Whitworth, A.~P.\ 2006, \aap, 450, 881 

\bibitem[Ikeda et al.(2008)]{Ikeda2008} Ikeda, H., et al.\ 2008, \aj, 135, 2323

\bibitem[Kessel-Deynet \& Burkert(2003)]{Kessel2003} Kessel-Deynet, O., \& Burkert, A.\ 2003, \mnras, 338, 545

\bibitem[Lefloch \& Lazareff(1994)]{LL94} Lefloch, B., \& Lazareff, B.\ 1994, \aap, 289, 559

\bibitem[Lefloch \& Lazareff(1995)]{LL95} Lefloch, B., \& Lazareff, B.\ 1995, \aap, 301, 522

\bibitem[Lefloch et al.(1997)]{Lefloch1997} Lefloch, B., Lazareff, B., \& Castets, A.\ 1997, \aap, 324, 249

\bibitem[Miao et al.(2009)]{Miao2009} Miao, J., White, G.~J., Thompson, M.~A., \& Nelson, R.~P.\ 2009, \apj, 692, 382

\bibitem[Monaghan(1992)]{Monaghan1992} Monaghan, J.~J.\ 1992, \araa, 30, 543 

\bibitem[Morgan et al.(2008)]{Morgan2008} Morgan, L.~K., Thompson, M.~A., Urquhart, J.~S., \& White, G.~J.\ 2008, \aap, 477, 557

\bibitem[Price (2007)]{Price2007} Price, D.~J.\ 2007, Publications of the Astronomical Society of Australia, 24, 159

\bibitem[Sandford et al.(1982)]{Sandford1982} Sandford, M.~T., II, Whitaker, R.~W., \& Klein, R.~I.\ 1982, \apj, 260, 183

\bibitem[Sugitani et al.(1999)]{Sugitani1999} Sugitani, K., Tamura, M., \& Ogura, K.\ 1999, Star Formation 1999, Proceedings of Star Formation 1999, held in Nagoya, Japan, June 21 - 25, 1999, Editor: T.~Nakamoto, Nobeyama Radio Observatory, p.~358-364, 358

\bibitem[Sugitani et al.(2000)]{Sugitani2000} Sugitani, K., Matsuo, H., Nakano, M., Tamura, M., \& Ogura, K.\ 2000, \aj, 119, 323

\end{thebibliography}
\end{document}